# A New Method For Digital Watermarking Based on Combination of DCT and PCA

Arash Saboori, S.Abolfazl Hosseini

*Abstract* — In the digital watermarking with DCT method, the watermark is located within a range of DCT coefficients of the cover image. In this paper to use the low-frequency band, a new method is proposed by using a combination of the DCT and PCA transform. The proposed method is compared to other DCT methods, our method is robust and keeps the quality of cover image, also increases capacity of the watermarking.

*Keywords* — Digital watermarking; DCT; low-frequency band; middle-frequency band; PCA transform.

## I. Introduction

THE Development of digital technology and the creation of wide digital networks such as internet, create Problems like unauthorized copying, fake ownership claims, because the digital documents can be copied or altered quickly and without any loss of quality. For avoiding these problems the watermarking system has been proposed [1]. In this system, a watermark that contains proprietary information by using a secret key hidden in the digital documents. The owners of document and authorized persons can extract the watermark Only by knowing the secret key and the watermarking's extraction algorithm [2]. The watermarking system has applications such as proof of ownership, authentication, checking the authenticity of the document, avoiding unauthorized copying [1]-[2]. The watermarking's mechanism is shown in Fig. 1.

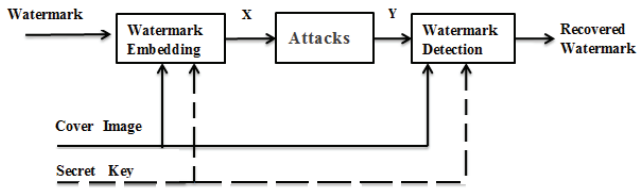

Fig. 1. The mechanism of a watermarking system [2].

The main characteristics of the watermarking system are determined by three factors: robustness, invisibility, capacity. These three components are in conflict always, the order of their importance depends on the application of the watermarking, as a matter of fact a compromise is between them [2]. The Watermarking techniques are categorized into spatial domain and frequency domain [3].

Arash Saboori is with Department of Communication, college of Electrical Engineering , Yadegar -e- Imam Khomeini (RAH) Branch, Islamic Azad University, Tehran, Iran (email: arash.saboori@yahoo.com).

S.Abolfazl Hosseini is with Department of Communication, college of Electrical Engineering, Yadegar-e-Imam Khomeini (RAH) Branch, Islamic Azad University,Tehran, Iran (email: abolfazl.hosseini@modares.ac.ir).

Although the spatial domain techniques need less hardware and have shorter execution time, but they cannot resist against various attacks. There are many techniques in frequency domains, i.e. DCT, DFT, DWT, etc. They are applied to the image, then a watermark will be added to the range of image frequency coefficient.

In this paper, we investigate various aspects of the watermarking with DCT method and a new scheme is proposed in order to increase its performance. Cox et.al [4] use DCT method for watermark embedding for the first time. They have believed the watermark should be placed in areas of the image that are important in term of human visual system (HVS), provided that the quality of cover image doesn't drop obviously and the watermark is quite invisible. In the DCT method, the middle frequency band is utilized usually according to a relative advantage compared to other bands [5], because the high-frequency band is quite fragile against most attacks and the low-frequency is not appropriate in term of invisibility of the watermark and the quality of the watermarked image while it is more resistant against some attacks (low-pass filtering and JPEG compression even with low quality factor) to the middle band. In this paper, in order to use the low-frequency band a new method is proposed based on a combination of the DCT and PCA technique. In our scheme the watermark is invisible while the quality of the cover image is maintained.

This paper consists of five parts. In the second and third part, the DCT and PCA technique will be reviewed. In the fourth section, the proposed scheme is presented, also evaluation of the proposed method and comparison with other methods is noted in this part. Finally, in part fifth, the results of the paper will be expressed generally.

## II. Discrete Cosine Transform

The two dimensional Discrete cosine transform (DCT) of an M ×N function $f(i,j)$ defined as follows [6]:

$$F(u,v) = \frac{2C(v)C(u)}{\sqrt{MN}} \sum_{i=0}^{M-1}\sum_{j=0}^{N-1} \cos\frac{(2i+1)u\pi}{2M} \cos\frac{(2j+1)v\pi}{2N} . f(i,j) \quad (1)$$

And inverse DCT given by

$$f(i,j) = \frac{2C(v)C(u)}{\sqrt{MN}} \sum_{i=0}^{M-1}\sum_{j=0}^{N-1} \cos\frac{(2i+1)u\pi}{2M} \cos\frac{(2j+1)v\pi}{2N} . F(i,j) \quad (2)$$

Where;

$$u = 0,1,\ldots, M\text{-}1, \quad v = 0, 1, \ldots, N-1, \quad C(\zeta)= \begin{cases} \frac{\sqrt{2}}{2} & if \ \zeta = 0 \\ 1 & otherwise \end{cases}$$

$F(u,v)$ is called DCT coefficients for function $f(i,j)$.

DCT coefficients correspond to a block of 8 × 8 of an image are changed according to Fig. 2 and are divided into three bands: the low-frequency band, the middle-frequency band and the high-frequency band [7].

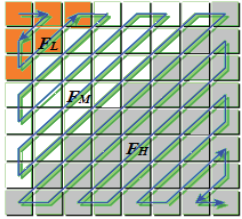

Fig. 2. Definition of DCT region based on Zig-Zag scanning pattern [7].

### III. PRINCIPAL COMPONENT ANALYSIS

Principal component analysis (PCA) is a technique for removing the linear dependence of the data. So, by that, given the reduced dimensions of the data [8].
Assuming matrix $X$ is a data set of size N×M, PCA transform is defined as [8]-[9] :

$$Y = AX \quad (3)$$

Where,
$\mathbf{X} = [\vec{X}_1, \vec{X}_2, \ldots, \vec{X}_k, \ldots, \vec{X}_N]^T$
$\vec{X}_k = [x_1, x_2, \ldots x_M]^T$ ; $(k=1,\ldots,N; \vec{X}_k \in R^M)$

The matrix $A$ contains the eigenvectors, $\vec{a}_i$, of the covariance matrix $X$, $C_X$, is computed as bellow steps [9]:

- The mean of matrix $\mathbf{X}$ is calculated as:

$$\vec{m} = \frac{1}{N}\sum_{k=1}^{N}\vec{X}_k \quad (4)$$

- The covariance matrix $\mathbf{X}$ is computed as formula (5).

$$\mathbf{C_X} = \frac{1}{N}\sum_{k=1}^{N}(\vec{X}_k - \vec{m})(\vec{X}_k - \vec{m})^T \quad (5)$$

- The eigenvectors, $\vec{a}_i$, of matrix $\mathbf{C_X}$, with corresponding eigenvalues, $\lambda_i$, satisfying formula (6).

$$\lambda_i \vec{a}_i = \mathbf{C_X} \vec{a}_i \quad (6)$$

Where $i=1, 2,\ldots M$, $\lambda_1 \geq \lambda_2 \geq \ldots \geq \lambda_M$ and $\vec{a}_i = [a_1, a_2, \ldots a_M]^T$, so the matrix $A$ is defined according to formula (7).

$$A = [\vec{a}_1, \vec{a}_2, \ldots \vec{a}_M] \quad (7)$$

So, the PCA through formula (3) maps the matrix data into principal component space, $Y$, the data has been completely non-correlated in this space [15].

### IV. THE PROPOSED SCHEME

In our scheme, the cover image due to the fact that nearby pixels together have a high correlation, is divided into non-overlapping blocks with the size of 8×8 and the DCT will be applied to each block separately, then the low-frequency band coefficients of block number k, are placed in the vector $\vec{X}_k$, Fig. 3.

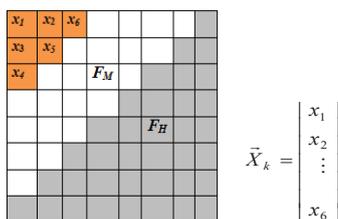

Fig. 3. The DCT region is used to embed watermark.

By repeating this process for all blocks, the data matrix $X$ will be formed, then the PCA transform will be applied to the data matrix $\mathbf{X}$. The first component of the PCA has the maximum energy concentration [10], in other words, it consists of common information of the DCT coefficients that are located in the low frequency band of a block. So the watermark is placed within it, because in this case the watermark is more resistant against attacks and it will be less affected. The mechanism of the proposed scheme is divided into two general sections, watermark embedding and watermark detection, as described in Fig. 1.

*A. Watermark Embedding Algorithm*

The algorithm which is used to embed a watermark on an image is given below:

1. The cover image is divided into blocks of 8×8.
2. The DCT will be applied to each block.
3. The low-frequency band coefficients are placed in the vector $\vec{X}_k$, then matrix data, $\mathbf{X}$, to be formed:
   $\mathbf{X} = [\vec{X}_1, \vec{X}_2, \ldots, \vec{X}_k \ldots \vec{X}_N]^T$; $(k=1,\ldots,N; \vec{X}_k \in R^M, M=6)$
4. The PCA will be applied to the matrix $\mathbf{X}$ and matrix $\mathbf{Y}$ is obtained; $\mathbf{Y} = [\vec{Y}_1, \vec{Y}_2, \ldots, \vec{Y}_k \ldots, \vec{Y}_N]$.
5. Convert the binary watermark logo into vector W
   $W = \{w_1, w_2, \ldots, w_N\}$ ; $w_i \in \{0,1\}, i=1,2,\ldots N$
6. Embed the watermark in the first component of PCA,
   $Y'_k = Y_k(i,1) + \alpha W_i \quad i=1,2,\ldots N \quad (8)$
   $\alpha$ based on a compromise between robustness and invisibility of watermarking is calculated.
7. The inverse PCA, IPCA, will be applied to matrix $\mathbf{Y}'$, so the matrix $\mathbf{X}'$ is restored.
8. The low frequency coefficients are placed in primary DCT blocks.
9. The inverse DCT, IDCT, will be applied to blacks and the watermarked image will be formed.

*B. Watermark Detection Algorithm*

The steps involved in the watermark detection algorithm are given below:

1. From the cover image and the watermarked image, matrix data, $\mathbf{X}$, will be formed separately.
   $\mathbf{X}$ = matrix data of cover image
   $\mathbf{X}'$ = matrix data of watermarked image
2. The PCA will be applied to matrix $\mathbf{X}$ and matrix $\mathbf{X}'$, so matrix $\mathbf{Y}$ and matrix $\mathbf{Y}'$ are obtained.
3. The watermark bits are extracted from the first component of the PCA as formula(9).

$$W_i = \frac{Y'_i - Y_i}{\alpha} \quad ; i=1,2,\ldots N \quad (9)$$

The block diagram of the proposed scheme is shown in Fig. 4. The size of the cover image is considered $m^2$ (m=8) times bigger than the watermark. The function $f$ is the intensity values in each block and the DCT is applied to each block, the low-frequency coefficient of block k are placed in $\vec{X}_k$, then the matrix data, $X$, will be formed. PC1 indicates the first principal component.

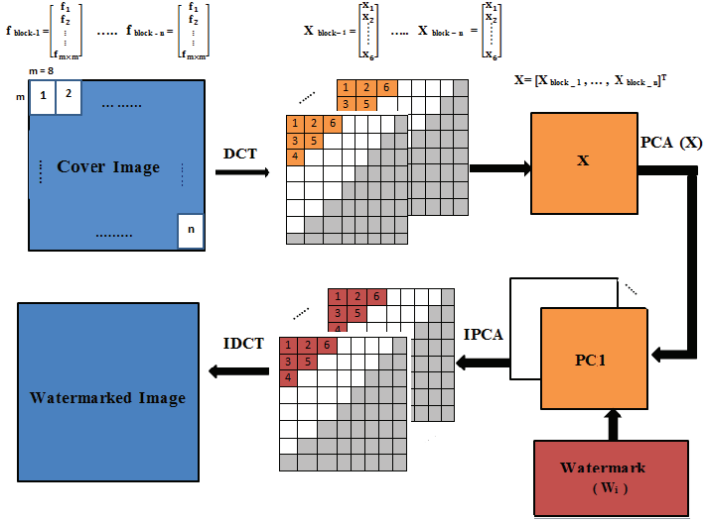

Fig. 4. The block diagram of the proposed scheme.

C. Evaluation Criteria

- Peak Signal to Noise Ratio (PSNR) is a criteria for evaluation of the image quality after the watermarking. Invisibility of a watermarking scheme can be described by PSNR. It is targeted to 40 dB [11]-[12].

$$mse = \frac{1}{mn}\sum_{i=1}^{m}\sum_{j=1}^{n}[I(i,j)-I'(i,j)]^2 \quad (10)$$

$$PSNR = 10\log_{10}(\frac{\max_{\forall(i,j)}(I(i,j))^2}{mse}) \quad (11)$$

Where, $I(i,j)$ = cover image; $I'(i,j)$ = watermarked image

- Normalized Correlation (NC) for comparison between the original watermark and the extracted watermark is used. It reflects to a robustness of watermarking scheme. NC is targeted to 1.0 [6].

$$NC = \frac{\sum_{i=1}^{m} W_i \times W_i'}{\sqrt{\sum_{i=1}^{m} W_i^2 \times \sum_{i=1}^{m} W_i'^2}} \quad (12)$$

Where, $w_i$=original watermark; $w_i'$=extracted watermark

- Capacity indicated the amount of bits can be embedded into the cover image.

D. Results and Discussion

In this simulation Lena image of size 512×512 is used as cover image and a binary image of size 64×64 is used as a watermark that are shown in Fig. 5. The proposed method is simulated by using MATLAB®

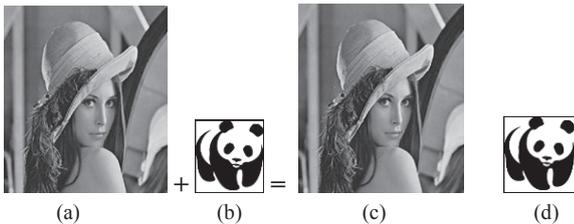

(a)  (b)  (c)  (d)

Fig. 5. (a) Cover image; (b) Watermark; (c) Watermarked image (PSNR=40. 1); (d) Extracted watermark (NC=1).

According to equation(8) and Table 1, the optimum value of coefficient α is determined based on compromise between PSNR and NC. For this simulation α=30 is intended.

TABLE 1: THE PROPOSED METHOD'S PSNR & NC WITHOUT ATTACKING FOR DIFFERENT COEFFICIENT α.

| α | 10 | 15 | 20 | 30 | 40 |
|---|---|---|---|---|---|
| PSNR(dB) | 47.8 | 44 | 42 | 40.1 | 36.56 |
| NC | 0.9888 | 0.9989 | 1 | 1 | 1 |

In Table 2 the watermarked images, go through some image processing attacks [13], such as JPEG compression, Gaussian filter, cropping, rotation and so on. PSNR between watermarked image and attacked image is calculated. Then, the watermark is extracted from the attacked image. Next, NC between the original watermark and the extracted watermark is calculated.

TABLE 2: SIMULATION RESULTS OF IMAGE PROCESSING ATTACKS.

| Attacks | (Q=70)JPEG | (Q=50)JPEG | (Q=20)JPEG | (Q=10)JPEG |
|---|---|---|---|---|
| Attacked image | | | | |
| PSNR(dB) | 36.90 | 35.86 | 33.11 | 30.65 |
| Extracted Watermarks | | | | |
| NC | 0.9837 | 0.9602 | 0.8143 | 0.5363 |
| Attacks | Gaussian noise (v=0.01) | Salt&pepper noise (v=0.01) | Sharpen | Rotate |
| Attacked image | | | | |
| PSNR(dB) | 19.79 | 25.14 | 26.64 | 23.16 |
| Extracted Watermarks | | | | |
| NC | 0.5270 | 0.7495 | 0.8309 | 0.3123 |
| Attacks | Median filter (3×3) | Average Filter (3×3) | Gaussian filter (3×3) | Crop |
| Attacked image | | | | |
| PSNR(dB) | 35.47 | 32.31 | 38.77 | 15.1 |
| Extracted Watermarks | | | | |
| NC | 0.9770 | 0.8623 | 1 | 0.7851 |

Table 3 shows proposed method's domain, size of the cover image and capacity in order to compare with counterparts [13]-[14]. Capacity is increased, because the amount of watermark bits in our methods is 64×64 while it is 32×32 for two counterparts.

TABLE 3: COMPARISON OF THE PROPOSED METHOD WITH OTHER METHODS IN TERM OF CAPACITY.

| Method | Domain | Cover image (Size) | Watermark (bits) |
|---|---|---|---|
| B. Reddy et al [13] | DCT | 512×512 | 32×32 |
| C. Mingzh et al [14] | DCT+DWT | 512×512 | 32×32 |
| Proposed method | DCT+PCA | 512×512 | 64×64 |

The experimental results are given in Table 4 in order to compare two methods [13]-[14]. It shows the NC and PSNR values after attacking by JPEG compression with different quality factors, low-pass filtering and some other attacks. The items have not been mentioned in the two counterparts is shown with a dark line in Table 4.

TABLE 4: COMPARISON OF THE PROPOSED METHOD WITH OTHER METHODS IN TERM OF ROBUSTNESS.

| Attacks | Proposed Scheme | | B. Reddy et al [13] | | C. Mingzh et al [14] | |
|---|---|---|---|---|---|---|
| | PSNR | NC | PSNR | NC | PSNR | NC |
| Without Attack | 40.1 | 1 | 38.89 | 1 | - | 1 |
| (Q=90)JPEG | 38.11 | 1 | 36.99 | - | - | 0.98 |
| (Q=70)JPEG | 36.90 | 0.9837 | 35.36 | 0.9866 | - | 0.97 |
| (Q=50)JPEG | 35.86 | 0.9602 | 34.40 | 0.8567 | - | 0.93 |
| (Q=30)JPEG | 34.46 | 0.8994 | 33.27 | 0.6655 | - | 0.89 |
| (Q=20)JPEG | 33.11 | 0.8143 | 32.29 | 0.5696 | - | 0.81 |
| (Q=10)JPEG | 30.65 | 0.5363 | 29.74 | 0.4074 | - | 0.37 |
| (Q=5)JPEG | 27.46 | 0.3456 | - | - | - | - |
| Gaussian noise (v=0.001) | 29.27 | 0.8921 | 31.8 | 0.9729 | - | 0.8895 |
| (3×3) median filter | 35.47 | 0.9770 | - | - | - | 0.9032 |
| (3×3) average filter | 32.31 | 0.8623 | - | - | - | 0.8432 |
| (3×3) Gaussian filter | 38.77 | 1 | 38.94 | 1 | - | 0.9117 |
| (5×5) Gaussian filter | 38.77 | 1 | 38.94 | 1 | - | - |
| (7×7) Gaussian filter | 38.77 | 1 | 38.94 | 1 | - | - |

The superiority of proposed method compared to other methods, in term of resistance against JPEG compression is shown in Fig. 6.

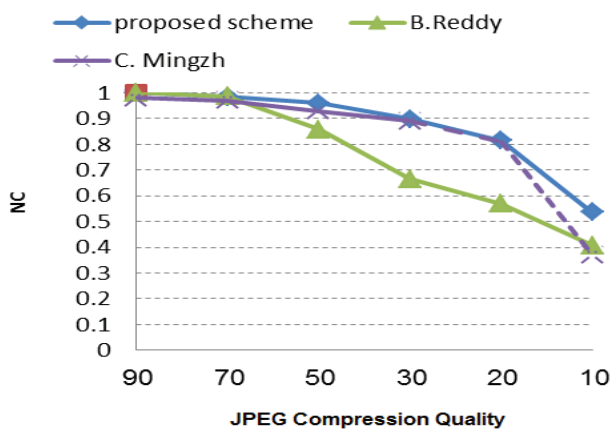

Fig. 6. Comparison proposed method with two counterparts [13]-[14] in term of JPEG compression.

The performance of the proposed scheme is better, because it is more resistant against attacks while the capacity of watermarking or watermark bits is increased. So specific feature of the proposed method is increasing the robustness and capacity relative to other methods.

## V. CONCLUSION

In this paper a new scheme is proposed in order to use of the low-frequency band of DCT coefficient based on a combination of DCT and PCA. Our scheme is robust against attacks, including low-pass filtering and JPEG compression with different quality factors. A specific feature of the proposed method is increasing the robustness and capacity relative to other methods that is shown in Tables 3 and 4.

REFERENCES

[1] Juergen Seitz, Tino Jahnke, "Digital Watermarking for Digital Media, "University of Cooperative Education Heidenheim, Germany, 2005, pp. 1–30.
[2] Michael Arnold, Martin Schmucker, Stephen D. Wolthusen, "Techniques and Applications of Digital Watermarking and Content Protection, " Artech House, 2003, pp. 55–69.
[3] Anu Bajaj, shagun, "Review on Watermarking Techniques: A Unique Approach for Digital Image Protection, " International Journal of Advanced Research in Computer Science and Software Engineering, volume 4, Issue 5, May 2014.
[4] Ingemar J. Cox, Senior Member, IEEE, Joe Kilian, F. Thomson Leighton, and Talal Shamoon, Member, IEEE, "Secure Spread Spectrum Watermarking for Multimedia," IEEE Transactions on Image Processing, Vol. 6, No. 12, December 1997.
[5] Tribhuwan Kumar Tewari, Vikas Saxena, "An Improved and Robust DCT based Digital Image Watermarking Scheme," International Journal of Computer Applications, Volume 3 – No.1, June 2010.
[6] Abduljabbar Shaamala, shahidan M. Abdullah and azizah A. Manaf, "Study of the effect DCT and DWT domains on the imperceptibility and robustness of Genetic watermarking," International Journal of Computer Science Issues, Vol. 8, Issue 5, No 2, September 2011.
[7] Angshumi Sarma and Amrita Ganguly, "Image Watermarking DCT-DWT Domain," IRNet Transactions on Electrical and Electronics Engineering (ITEEE), Vol-1, Iss-2, 2012.
[8] I.T. Jolliffe, Principal Component Analysis, 2nd Edn., New York: Springer-Verlag (2002).
[9] Craig Rodarmel, Jie Shan, "Principal Component Analysis for Hyperspectral Image Classification," Surveying and Land Information Systems, Vol. 62, No. 2, 2002, pp.115-000.
[10] K. S. Karpe, Dr. S. K. Shah, Dr. P. Mukherji, "Hybrid Digital Video Watermarking based on DWT-PCA," International Journal of Advance Research in Computer Science and Management Studies, Volume 1, Issue 5, October 2013.
[11] S.Abolfazl Hosseini, Besharat Rezaei Shookooh, Saeid Shahhosseini, Shahriar Beizaee, " Speeding up Fractal Image De-Compression," International Conference on Computer Applications and Industrial Industrial Electronics, 2010.
[12] Saeid Shahhosseini, S. Abolfazl Hosseini, "A Visual Base Image Coding Using Unequal Error Protection in High Error Rate Channels," International Conference on Computer Applications and Industrial Electronics , December 5-7, 2010.
[13] B. Rakesh Reddy, G.Nirmala, Dr. A. Yesu Babu, "Robust and Oblivious Watermarking Scheme Based on DCT," IJCST Vo l . 4, Issue Spl - 4, Oct - Dec 2013.
[14] Cheng Mingzhi, Li Yan, Zhou Yajian and Lei Min, "A Combined DWT and DCT Watermarking," Journal Of Multimedia, Vol. 8, No. 3, June 2013.
[15] Hamid Reza Shahdoosti, Hassan Ghassemian, "Spatial PCA as A New Method For Image Fusion," The 16th CSI International Symposium on Artificial Intelligence and Signal Processing (AISP 2012).